\documentclass[preprint,showpacs,preprintnumbers,amsmath,amssymb]{revtex4}

\usepackage{graphicx}
\usepackage{dcolumn}
\usepackage{bm}

\begin{document}

\preprint{}

\title{An improved exact upper bound of $\frac{22}{35}$ on the Hilbert-Schmidt separability probability of real two-qubit systems}
\author{Paul B. Slater}%
\email{slater@kitp.ucsb.edu}
\affiliation{%
ISBER, University of California, Santa Barbara, CA 93106\\
}%
\date{\today}

\begin{abstract}
We seek to derive the probability--expressed in terms of the Hilbert-Schmidt (Euclidean or flat) metric--that a generic (nine-dimensional) real two-qubit system is separable, by implementing the well-known Peres-Horodecki test on the partial transposes (PTs) of the associated $4 \times 4$ density matrices ($\rho$). But the full implementation of the test--requiring that the determinant of the PT be nonnegative for separability to 
hold--appears to be, at least presently, computationally intractable.
So, we have previously implemented--using the auxiliary concept of a {\it diagonal-entry-parameterized separability function} (DESF)--the weaker implied test of nonnegativity   of the six $2 \times 2$ principal minors of the PT. This yielded an exact 
upper bound on the separability probability of 
$\frac{1024}{135 \pi^2} \approx 0.76854$. Here, we extend this line of work by requiring that the four $3 \times 3$ principal minors of the PT be nonnegative, giving us an improved/reduced upper bound of $\frac{22}{35} \approx 0.628571$. Numerical simulations--as opposed to exact symbolic 
calculations--indicate, on the other hand, that the true probability is certainly less than $\frac{1}{2}$. Our combined analyses lead us to suggest a possible form for the true DESF, yielding a separability probability of 
$\frac{29}{64} \approx 0.453125$, while the best exact {\it lower} bound established so far is $\frac{6928-2205 \pi }{2^{9/2}} 
\approx 0.0348338$. 
\end{abstract}

\pacs{Valid PACS 03.67.Mn, 02.10.Ud, 02.30.Cj, 02.40.Ft, 02.40.Ky}
\keywords{two qubits, separability probabilities, separability functions, Peres-Horodecki conditions, partial transpose, real density matrices, matrix minors, nonnegativity, quasi-Monte Carlo, numerical integration, Hilbert-Schmidt metric}

\maketitle

\begin{figure}
\includegraphics{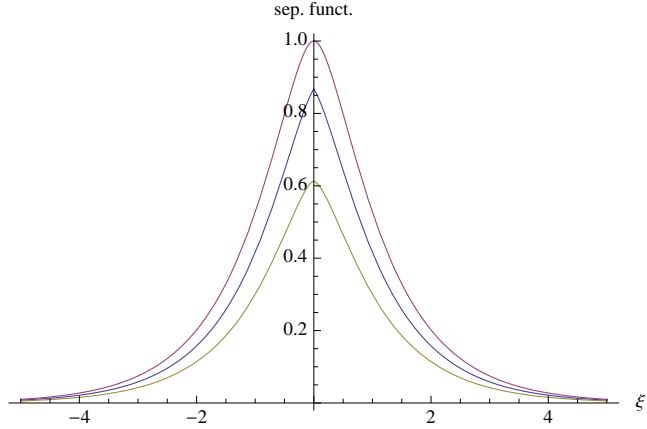}
\caption{\label{fig:threecurves}Three forms of diagonal-entry-parameterized separability functions (DESFs)}
\end{figure}

We direct the reader's attention to Fig.~\ref{fig:threecurves}, which depicts various forms of "diagonal-entry-parameterized separability functions" (DESF's) \cite{slaterPRA2,slater833}--as opposed to "eigenvalue-parameterized separability functions (ESFs) \cite{maxconcur4,maxconcur2,JMP2008}--that we will employ here to obtain estimates and simple exact upper bounds on the Hilbert-Schmidt (HS) probability that a generic (nine-dimensional) real two-qubit system is separable.
{\.Z}yczkowski, Horodecki, Sanpera and Lewenstein, in a much-cited article \cite{ZHSL},  have given "philosophical", "practical" and "physical" reasons for studying "separability probabilities". We have examined the associated problems which arise, using the volume elements of several metrics of interest as measures on the quantum states, in various numerical and theoretical studies \cite{slaterA,slaterC,slaterJGP,slaterPRA,pbsCanosa,slater833,JMP2008,ratios}. 

\begin{figure}
\includegraphics{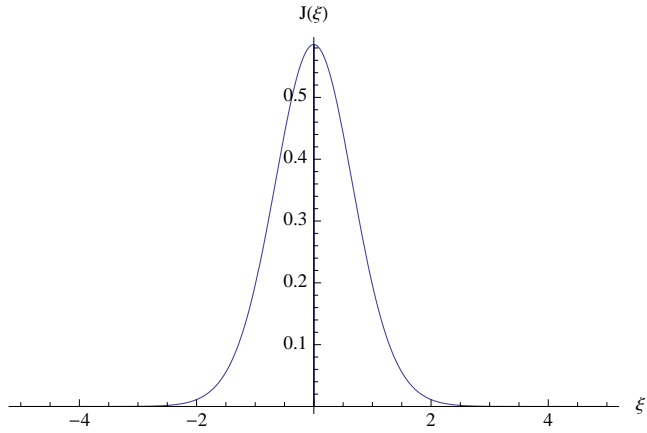}
\caption{\label{fig:JacobianMinors}Jacobian (\ref{jacobian}), which when multiplied by a separability function and integrated over $\xi \in [-\infty,\infty]$, yields the associated Hilbert-Schmidt separability probability}
\end{figure}

The subordinate of the three curves in 
Fig.~\ref{fig:threecurves}--derived using an extensive quasi-Monte Carlo (Tezuka-Faure \cite{giray1,tezuka}) six-dimensional numerical integration procedure--provides an estimate of the true, but so-far not exactly-determined DESF. The dominant of the three curves--readily obtainable from results already reported in 
\cite[sec. VII]{slaterPRA2}--has the form
\begin{equation} \label{dominantcurve}
S_{dom}(\xi)=\begin{cases}
\frac{1}{2} e^{-3 \xi } \left(3 e^{2 \xi }-1\right) & \xi >0 \\
 -\frac{1}{2} e^{\xi } \left(e^{2 \xi }-3\right) & \xi <0
\end{cases}.
\end{equation}
The intermediate of the three curves, which we first report here, has the same--differing only in constants--functional form
\begin{equation} \label{intermediatecurve}
S_{int}(\xi)= \begin{cases}
 \frac{9 \pi^2}{2048}  e^{-3 \xi } \left(27 e^{2 \xi }-7\right) & \xi >0
   \\
 -\frac{9 \pi^2}{2048}  e^{\xi } \left(7 e^{2 \xi }-27\right) & \xi <0
\end{cases}.
\end{equation}
With each of these three curves we can obtain an associated estimate or upper bound on
the desired HS separability probability ($P^{HS}_{sep/real}$). This is accomplished by integrating over $\xi \in [-\infty,\infty]$ the product of the corresponding curve with the function (Fig.~\ref{fig:JacobianMinors}) (based on the jacobian of a 
coordinate transformation, to be described below)
\begin{equation} \label{jacobian}
J(\xi)= \frac{64 \text{csch}^9(\xi ) (-160 \sinh (2 \xi )-25 \sinh (4 \xi )+12
   \xi  (16 \cosh (2 \xi )+\cosh (4 \xi )+18))}{27 \pi ^2},
\end{equation}
that is,
\begin{equation}
P^{HS}_{sep/real}=\int^{\infty}_{-\infty} S(\xi) J(\xi) \mbox{d} \xi.
\end{equation}
Proceeding, thusly, we obtain an upper bound on the HS separability probability of 
$\frac{1024}{135 \pi^2} \approx 0.76854$ based on the dominant of the three curves, the titular 
$\frac{22}{35} \approx 0.628571$ using the intermediate curve, and an estimate of 0.4528427 for the true probability with the subordinate, numerically-derived curve. (From our work in \cite[eq. (25)]{maxconcur2}, we already know that the HS probability of a generic real two-qubit system being {\it absolutely} separable--that is not entanglable by any unitary transformation--is $\frac{6928-2205 \pi }{2^{9/2}} 
\approx 0.0348338$, which then serves as a {\it lower} bound on the corresponding HS [absolute {\it plus} nonabsolute] separability probability itself (cf. \cite{sepsize1} \cite[eq. (29)]{maxconcur2}).)

The variable $\xi$ used in the above presentation is the logarithm of the square root of the 
ratio of
the product of the 11- and 44-entry of the associated real $4 \times 4$ density matrix 
($\rho$) to the product of the 22- and 33-entries, that is 
\begin{equation} \label{xidefinition}
\xi =\log{\sqrt{\frac{\rho_{11} \rho_{44}}{\rho_{22} \rho_{33}}}}= \frac{1}{2} \log{\frac{\rho_{11} \rho_{44}}{\rho_{22} \rho_{33}}}.
\end{equation}
(In our previous studies \cite{slaterPRA2,slater833}, we have employed the alternative variables, 
$\nu =\frac{\rho_{11} \rho_{44}}{\rho_{22} \rho_{33}}$ and 
$\mu =\sqrt{\frac{\rho_{11} \rho_{44}}{\rho_{22} \rho_{33}}}$, but now switch to the [more symmetric] form (\ref{xidefinition}). Importantly, only the "cross-product ratio of diagonal entries is needed in our parameterization to test for separability, and not the individual entries themselves.)
The jacobian (\ref{jacobian}) used in our calculations is obtained by the transformation
of one of the diagonal entries, say, $\rho_{33}$, to $\xi$ and integrating the Hilbert-Schmidt 
(Lebesgue) volume element (of course, $\rho_{44}=1-\rho_{11}-\rho_{22}-\rho_{33}$) \cite[p. 13646]{andai}
\begin{equation}
\mbox{d} V_{HS}= (\rho_{11} \rho_{22} \rho_{33} \rho_{44})^{\frac{3 \beta}{2}} 
\mbox{d} \rho_{11} \mbox{d} \rho_{22} \mbox{d} \rho_{33}, \hspace{.25in} \beta=1
\end{equation}
over $\rho_{11}$ and $\rho_{22}$ and normalizing the result. (To obtain the corresponding HS volume elements for the {\it complex} $4 \times 4$ density matrices, one must employ--conforming to a pattern familiar from random matrix theory--$\beta=2$, and 
$\beta=4$ in the {\it quaternionic} case (cf. \cite{szHS}).)

\begin{figure}
\includegraphics{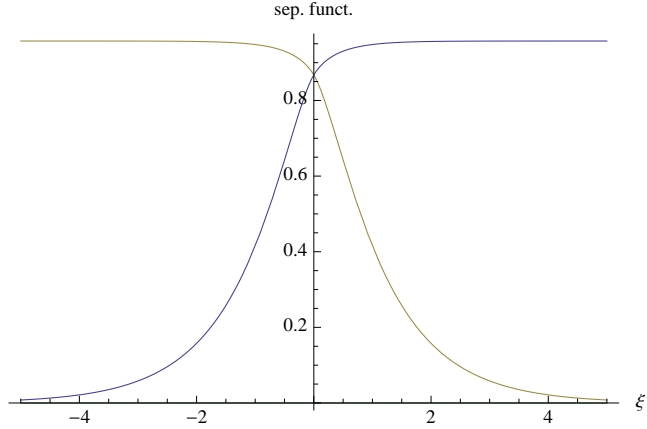}
\caption{\label{fig:SymmetricCurves}The two distinct (red and blue) separability functions obtained from the four $3 \times 3$ principal minors, the envelope of which defines the intermediate curve in Fig.~\ref{fig:threecurves}. The intercept, as well as the point of intersection of the two curves, is at 
$\frac{45 \pi^2}{512} \approx 0.867446$.}
\end{figure}

\begin{figure}
\includegraphics{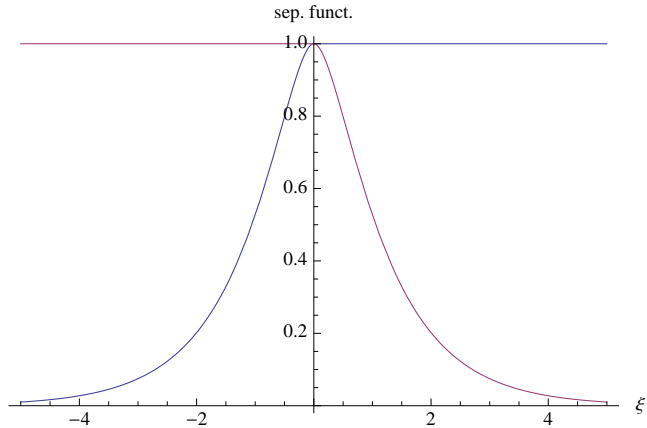}
\caption{\label{fig:SymmetricCurves2}The two distinct (red and blue) separability functions obtained from the six $2 \times 2$ principal minors, the envelope of which defines the dominant curve in Fig.~\ref{fig:threecurves}}
\end{figure}

The use of the celebrated Peres-Horodecki separability test \cite{asher,michal} is central to our analyses. Ideally, we would be able to require that the determinant of the partial transpose of 
$\rho$ be nonnegative to guarantee separability \cite{augusiak,ver}. However, this has so far proved to be too computationally demanding a (fourth-degree, high-dimensional) task for us to enforce (cf. \cite[eq. (7)]{slaterPRA2}). But, in
\cite{slater833}, we did succeed in implementing the weaker implied test that all the six $2 \times 2$ principal minors of the partial transpose of $\rho$ be nonnegative, giving us the dominant curve in Fig.~\ref{fig:threecurves}. (Actually, only two of the minors differ nontrivially from the analogous set of minors of 
$\rho$ itself.) To derive the sharper intermediate curve here, we extended this approach to the four $3 \times 3$ principal minors. Actually, we found that these four minors fell into two pairs of identical results. Further, one of the set of results 
\begin{equation} \label{oneresult}
S_{3 \times 3}(\xi)= \begin{cases}
\frac{9 \pi ^2 e^{-3 \xi } \left(27 e^{2 \xi }-7\right)}{2048} & \xi >0
   \\
 \frac{3 \pi  e^{-3 \xi } \left(e^{\xi } \sqrt{1-e^{2 \xi }} \left(37
   e^{2 \xi }+2 e^{4 \xi }+21\right)+3 \left(27 e^{2 \xi }-7\right) \sin
   ^{-1}\left(e^{\xi }\right)\right)}{1024} & \xi <0
\end{cases}
\end{equation}
could be obtained from the other set by the transformation $\xi \rightarrow -\xi$. This curve (\ref{oneresult}) and its reflection around $\xi=0$ are shown in Fig.~\ref{fig:SymmetricCurves}. The intermediate curve (\ref{intermediatecurve}) in Fig.~\ref{fig:threecurves}, first reported here, was constructed by joining the sharper segments of these two curves over the two half-axes. (A parallel strategy had been pursued with the $2 \times 2$ minors.)
The comparable results to (\ref{oneresult}) and 
Fig.~\ref{fig:SymmetricCurves} for the $2 \times 2$ minors investigation \cite{slater833}
are
\begin{equation}
 S_{2 \times 2}(\xi)=\begin{cases}
 e^{-2 \xi } (2 \sinh (\xi )+\cosh (\xi )) & \xi >0 \\
1 & \xi <0 
\end{cases}
\end{equation}
and Fig.~\ref{fig:SymmetricCurves2}.

For the intermediate curve in Fig.~\ref{fig:threecurves} we have the nontrivial $y$-axis intercept of $\frac{45 \pi^2}{512} \approx 0.867446$ 
(the intercept for the dominant curve being simply 1), while the estimate of the true intercept using the numerically-generated curve is 
0.612243, quite close to our previously conjectured value of 
$\frac{135 \pi^2}{2176} \approx 0.612315$ \cite{slater833}.

In obtaining our several results, we used the "Bloore/correlation"  parameterization of density matrices 
\cite{bloore,joe} and accompanying ranges of integration--generated by the cylindrical algebraic decomposition procedure \cite{cylindrical,strzebonski}, implementing the requirement that $\rho$ be nonnegative definite--presented in \cite[eqs. (3)-(5)]{slaterPRA2}. The  computational tractability of utilizing the $3 \times 3$ principal minors of the partial transpose in this coordinate frame appeared to stem from the fact that each of these four quantities only contains  
{\it three} of the six off-diagonal variables ($z_{ij}$) employed in the full parameterization (each set of three variables, additionally and conveniently, sharing a common row/column subscript). (The nine-dimensional convex set of real two-qubit density matrices is parameterized by six off-diagonal--$z_{ij}=\frac{\rho_{ij}}{\sqrt{\rho_{ii} \rho_{jj}}}$--and three diagonal variables--$\rho_{ii}$.) Integrating out the three variables not present in the constraint simply leaves us with a constrained (boolean) integration over the cube $[-1,1]^3$, as indicated in \cite[eq. (3)]{slaterPRA2}. We appropriately permuted the subscripts in the indicated coordinate system, 
so that we could study all four of the minors (thus, finding that they fell into two equal sets). Of course, such a simplifying integration strategy is not available for the determinant of the partial transpose itself, which contains all the six off-diagonal variables ($z_{ij}$), rather than simply three.

Each of the constrained integrations we have utilized so far, has used as it constraint the nonnegativity of a {\it single} $2 \times 2$ or 
$3 \times 3$ principal minor of the partial transpose of $\rho$. (However, we have been able above to couple and complement multiple results by taking the 
sharper/tighter bounds over the half-axes provided by individual outcomes.) We have, to this point, been unable--using either the (Bloore \cite{bloore}) density-matrix parameterization presented in \cite{slaterPRA2} or the interesting partial-correlation parameterization indicated in \cite{joe}--to perform constrained integrations in which two or more $2 \times 2$ or $3 \times 3$ minors (and {\it a fortiori} the determinant) are required to be {\it simultaneously} nonnegative. 
(It, then, remains an open question whether or not being able to do so would simply lead to the dominant and intermediate curves given already in Fig.~\ref{fig:threecurves} and by 
(\ref{dominantcurve}) and (\ref{intermediatecurve})).

We can, however, rather convincingly--but in a somewhat heuristic manner--reduce the derived upper bound on the HS separability probability of generic real two-qubit systems from $\frac{22}{35} \approx 0.628571$ to 0.576219 by using a new curve--having a $y$-intercept of $(\frac{45 \pi^2}{512})^2 \approx 0.752462$ as a DESF. This curve is obtained by taking the {\it product} of the two curves displayed in Fig.~\ref{fig:SymmetricCurves} (that is, the product of the function (\ref{intermediatecurve}) with its reflection about $\xi=0$). A plot of the result shows that
it is both subordinate to the intermediate curve in Fig.~\ref{fig:threecurves}, as is obvious it must be, but also clearly dominates the numerically-generated curve there, which is an estimate of the true DESF. (Since each of the two curves in Fig.~\ref{fig:SymmetricCurves2} is simply unity over a half-axis, a parallel strategy in the $2 \times 2$ minors analysis can, of course, yield no nontrivial upper-bound reduction from 
$\frac{1024}{135 \pi^2} \approx 0.76854$.)

The "twofold-ratio" theorem of Szarek, Bengtsson and 
{\.Z}yczkowski \cite{sbz}--motivated by the numerical results reported in \cite{slaterPRA}--allows us to immediately obtain exact upper bounds, as well, on the HS separability probability for generic (eight-dimensional) real 
{\it minimally-degenerate} real two-qubit systems (boundary states having a single eigenvalue zero). These upper bounds would, then, be {\it one-half} those applicable to the nondegenerate case--that is, $\frac{512}{135 \pi^2} \approx 0.38427$ and $\frac{11}{35} 
\approx 0.314286$. Further, we can, using the results of our numerical study, similarly obtain an induced estimate, 0.226421, of the true probability.

The two sets of derived functions (\ref{dominantcurve}) and (\ref{intermediatecurve}), based 
respectively on the $2 \times 2$ and $3 \times 3$ minors have the same functional forms, but with differing sets of constants ($\{1,2,3,1\}$ {\it vs.} $\{9,2048=2^{11},27,7\}$). It seems natural, then, to conjecture that the true separability function--which must be based on the determinant of the partial transpose \cite[eq. (7)]{slaterPRA2} \cite{augusiak,ver}, that is, 
the single $4 \times 4$ minor--will also adhere to the same functional form, but with a different set of constants.

\begin{figure}
\includegraphics{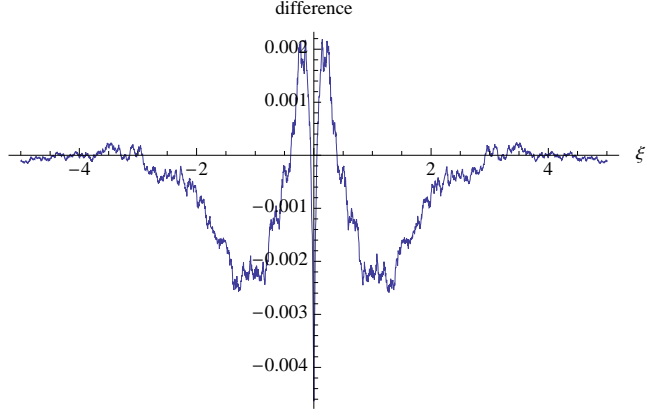}
\caption{\label{fig:differences}The difference between the numerically-generated subordinate function in Fig.~\ref{fig:differences} and a  suggested possibly true separability function (\ref{conjecture}), giving a separability probability of $\frac{29}{64} \approx 0.453125$}
\end{figure}

\begin{figure}
\includegraphics{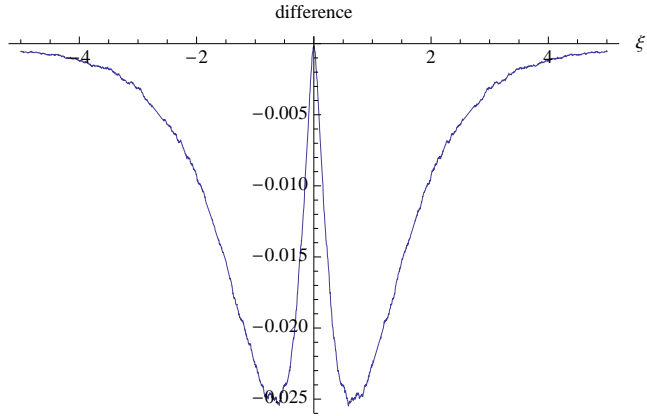}
\caption{\label{fig:differences2}The difference between the numerically-generated subordinate function in Fig.~\ref{fig:differences} and the previously conjectured true separability function (\ref{conjecture2}), giving a separability probability of $\frac{8}{17} \approx 0.470588$}
\end{figure}

In fact, pursuing this line of thought, we have found that the function 
\begin{equation} \label{conjecture}
S_{conjecture}(\xi)=\begin{cases}
\frac{315 e^{-3 \xi } \left(-5+18 e^{2 \xi }\right) \pi ^2}{2^{16}} & \xi
   >0 \\
 -\frac{315 e^{\xi } \left(-18+5 e^{2 \xi }\right) \pi ^2}{2^{16}} & \xi
   <0 
 \end{cases}
\end{equation}
fits (Fig.~\ref{fig:differences}) the numerically-generated subordinate curve in Fig.~\ref{fig:threecurves} quite well, yielding an HS separability probability of 
$\frac{29}{64} \equiv \frac{29}{2^6} \approx  0.453125$, and a $y$-intercept of $\frac{4095 \pi^2}{2^{16}} \approx 0.6167$.  (Then, by the twofold-ratio theorem \cite{sbz}, the HS separability probability of the minimally-degenerate (boundary) states would be $\frac{29}{128} \equiv \frac{29}{2^7}  \approx 0.226563$. Also, we have been able to find a number of other curves, adhering to this same general structure, fitting the subordinate curve in Fig.~\ref{fig:threecurves} equally as well, and again  yielding $\frac{29}{64}$ as a separability probability, in addition to 
well-fitting curves yielding somewhat less simple fractions--such 
as $\frac{163}{360} \approx 0.452778, \frac{367}{810} \approx 0.453086$ and
$\frac{428}{945} \approx 0.45291$.)
We are obligated, however, to note that in 
\cite[sec. IX.A]{slater833} we had advanced--based on somewhat different considerations than here--the hypothesis that this probability is 
$\frac{8}{17} \approx 0.470588$, with an associated DESF equal to 
\begin{equation} \label{conjecture2}
S_{previous}(\xi)=\begin{cases}
\frac{135 e^{-3 \xi } \left(-1+3 e^{2 \xi }\right) \pi ^2}{2^8 \cdot 17} & \xi
   >0 \\
 -\frac{135 e^{\xi } \left(-3+e^{2 \xi }\right) \pi ^2}{2^8 \cdot 17} & \xi <0 
\end{cases}.
\end{equation}
(However, our best numerical estimate at that point was 0.4538838 \cite[sec. V.A.2]{slaterPRA2} \cite[sec. IX.A]{slater833}, rather close to our current-study estimate of 0.4528427. By computing standard errors of the mean, we can establish a [$\approx 95\%$] confidence range--four standard deviations wide--for this estimate of $(0.451634, 0.454051)$--that does contain 
$\frac{29}{64} \approx 0.453125$.
A comparable plot (Fig.~\ref{fig:differences2}) to 
Fig.~\ref{fig:differences} shows (\ref{conjecture2}) to provide a considerably poorer fit.)

One might further speculate--in line with random matrix theory and our previous analyses \cite{slater833}--that the DESF for the generic (15-dimensional) {\it complex} two-qubit systems is proportional to the {\it square} of (\ref{conjecture}). If the constant of proportionality were simply taken to equal unity, the associated HS separability probability would be
$\frac{30660525 \pi ^4}{11811160064}= \frac{3^5 \cdot 5^2 \cdot 7^2 \cdot 103 \pi ^4}{ 2^{30} \cdot 11} \approx 0.252864$, rather close to the value 
$\frac{8}{33} \approx 0.242424$ conjectured, for a number of reasons, in \cite[sec. IX.B]{slater833}.

\begin{acknowledgments}
I would like to express appreciation to the Kavli Institute for Theoretical
Physics (KITP)
for computational support in this research.
\end{acknowledgments}

\bibliography{Minors}

\begin{thebibliography}{26}
\expandafter\ifx\csname natexlab\endcsname\relax\def\natexlab#1{#1}\fi
\expandafter\ifx\csname bibnamefont\endcsname\relax
  \def\bibnamefont#1{#1}\fi
\expandafter\ifx\csname bibfnamefont\endcsname\relax
  \def\bibfnamefont#1{#1}\fi
\expandafter\ifx\csname citenamefont\endcsname\relax
  \def\citenamefont#1{#1}\fi
\expandafter\ifx\csname url\endcsname\relax
  \def\url#1{\texttt{#1}}\fi
\expandafter\ifx\csname urlprefix\endcsname\relax\def\urlprefix{URL }\fi
\providecommand{\bibinfo}[2]{#2}
\providecommand{\eprint}[2][]{\url{#2}}

\bibitem[{\citenamefont{Slater}(2007{\natexlab{a}})}]{slaterPRA2}
\bibinfo{author}{\bibfnamefont{P.~B.} \bibnamefont{Slater}},
  \bibinfo{journal}{Phys. Rev. A} \textbf{\bibinfo{volume}{75}},
  \bibinfo{pages}{032326} (\bibinfo{year}{2007}{\natexlab{a}}).

\bibitem[{\citenamefont{Slater}(2007{\natexlab{b}})}]{slater833}
\bibinfo{author}{\bibfnamefont{P.~B.} \bibnamefont{Slater}},
  \bibinfo{journal}{J. Phys. A} \textbf{\bibinfo{volume}{40}},
  \bibinfo{pages}{14279} (\bibinfo{year}{2007}{\natexlab{b}}).

\bibitem[{\citenamefont{Slater}(2008)}]{maxconcur4}
\bibinfo{author}{\bibfnamefont{P.~B.} \bibnamefont{Slater}},
  \bibinfo{journal}{J. Phys. A} \textbf{\bibinfo{volume}{41}},
  \bibinfo{pages}{505303} (\bibinfo{year}{2008}).

\bibitem[{\citenamefont{Slater}({\natexlab{a}})}]{maxconcur2}
\bibinfo{author}{\bibfnamefont{P.~B.} \bibnamefont{Slater}},
  \eprint{arXiv:0905.0161 (to appear in J. Phys. A)}.

\bibitem[{\citenamefont{Slater}(2009)}]{JMP2008}
\bibinfo{author}{\bibfnamefont{P.~B.} \bibnamefont{Slater}},
  \bibinfo{journal}{J. Geom. Phys.} \textbf{\bibinfo{volume}{59}},
  \bibinfo{pages}{17} (\bibinfo{year}{2009}).

\bibitem[{\citenamefont{{\.Z}yczkowski
  et~al.}(1998)\citenamefont{{\.Z}yczkowski, Horodecki, Sanpera, and
  Lewenstein}}]{ZHSL}
\bibinfo{author}{\bibfnamefont{K.}~\bibnamefont{{\.Z}yczkowski}},
  \bibinfo{author}{\bibfnamefont{P.}~\bibnamefont{Horodecki}},
  \bibinfo{author}{\bibfnamefont{A.}~\bibnamefont{Sanpera}}, \bibnamefont{and}
  \bibinfo{author}{\bibfnamefont{M.}~\bibnamefont{Lewenstein}},
  \bibinfo{journal}{Phys. Rev. A} \textbf{\bibinfo{volume}{58}},
  \bibinfo{pages}{883} (\bibinfo{year}{1998}).

\bibitem[{\citenamefont{Slater}(1999)}]{slaterA}
\bibinfo{author}{\bibfnamefont{P.~B.} \bibnamefont{Slater}},
  \bibinfo{journal}{J. Phys. A} \textbf{\bibinfo{volume}{32}},
  \bibinfo{pages}{5261} (\bibinfo{year}{1999}).

\bibitem[{\citenamefont{Slater}(2000)}]{slaterC}
\bibinfo{author}{\bibfnamefont{P.~B.} \bibnamefont{Slater}},
  \bibinfo{journal}{Euro. Phys. J. B} \textbf{\bibinfo{volume}{17}},
  \bibinfo{pages}{471} (\bibinfo{year}{2000}).

\bibitem[{\citenamefont{Slater}(2005{\natexlab{a}})}]{slaterJGP}
\bibinfo{author}{\bibfnamefont{P.~B.} \bibnamefont{Slater}},
  \bibinfo{journal}{J. Geom. Phys.} \textbf{\bibinfo{volume}{53}},
  \bibinfo{pages}{74} (\bibinfo{year}{2005}{\natexlab{a}}).

\bibitem[{\citenamefont{Slater}(2005{\natexlab{b}})}]{slaterPRA}
\bibinfo{author}{\bibfnamefont{P.~B.} \bibnamefont{Slater}},
  \bibinfo{journal}{Phys. Rev. A} \textbf{\bibinfo{volume}{71}},
  \bibinfo{pages}{052319} (\bibinfo{year}{2005}{\natexlab{b}}).

\bibitem[{\citenamefont{Slater}(2006)}]{pbsCanosa}
\bibinfo{author}{\bibfnamefont{P.~B.} \bibnamefont{Slater}},
  \bibinfo{journal}{J. Phys. A} \textbf{\bibinfo{volume}{39}},
  \bibinfo{pages}{913} (\bibinfo{year}{2006}).

\bibitem[{\citenamefont{Slater}({\natexlab{b}})}]{ratios}
\bibinfo{author}{\bibfnamefont{P.~B.} \bibnamefont{Slater}},
  \eprint{arXiv:0905.0161}.

\bibitem[{\citenamefont{{\"O}kten}(1999)}]{giray1}
\bibinfo{author}{\bibfnamefont{G.}~\bibnamefont{{\"O}kten}},
  \bibinfo{journal}{MATHEMATICA in Educ. Res.} \textbf{\bibinfo{volume}{8}},
  \bibinfo{pages}{52} (\bibinfo{year}{1999}).

\bibitem[{\citenamefont{Faure and Tezuka}(2002)}]{tezuka}
\bibinfo{author}{\bibfnamefont{H.}~\bibnamefont{Faure}} \bibnamefont{and}
  \bibinfo{author}{\bibfnamefont{S.}~\bibnamefont{Tezuka}}, in
  \emph{\bibinfo{booktitle}{Monte Carlo and Quasi-Monte Carlo Methods 2000
  (Hong Kong)}}, edited by \bibinfo{editor}{\bibfnamefont{K.~T.}
  \bibnamefont{Tang}}, \bibinfo{editor}{\bibfnamefont{F.~J.}
  \bibnamefont{Hickernell}}, \bibnamefont{and}
  \bibinfo{editor}{\bibfnamefont{H.}~\bibnamefont{Niederreiter}}
  (\bibinfo{publisher}{Springer}, \bibinfo{address}{Berlin},
  \bibinfo{year}{2002}), p. \bibinfo{pages}{242}.

\bibitem[{\citenamefont{Gurvits and Barnum}(2002)}]{sepsize1}
\bibinfo{author}{\bibfnamefont{L.}~\bibnamefont{Gurvits}} \bibnamefont{and}
  \bibinfo{author}{\bibfnamefont{H.}~\bibnamefont{Barnum}},
  \bibinfo{journal}{Phys.Rev. A} \textbf{\bibinfo{volume}{66}},
  \bibinfo{pages}{062311} (\bibinfo{year}{2002}).

\bibitem[{\citenamefont{Andai}(2006)}]{andai}
\bibinfo{author}{\bibfnamefont{A.}~\bibnamefont{Andai}}, \bibinfo{journal}{J.
  Phys. A} \textbf{\bibinfo{volume}{39}}, \bibinfo{pages}{13641}
  (\bibinfo{year}{2006}).

\bibitem[{\citenamefont{{\.Z}yczkowski and Sommers}(2003)}]{szHS}
\bibinfo{author}{\bibfnamefont{K.}~\bibnamefont{{\.Z}yczkowski}}
  \bibnamefont{and} \bibinfo{author}{\bibfnamefont{H.-J.}
  \bibnamefont{Sommers}}, \bibinfo{journal}{J. Phys. A}
  \textbf{\bibinfo{volume}{36}}, \bibinfo{pages}{10115} (\bibinfo{year}{2003}).

\bibitem[{\citenamefont{Peres}(1996)}]{asher}
\bibinfo{author}{\bibfnamefont{A.}~\bibnamefont{Peres}},
  \bibinfo{journal}{Phys. Rev. Lett.} \textbf{\bibinfo{volume}{77}},
  \bibinfo{pages}{1413} (\bibinfo{year}{1996}).

\bibitem[{\citenamefont{Horodecki et~al.}(1996)\citenamefont{Horodecki,
  Horodecki, and Horodecki}}]{michal}
\bibinfo{author}{\bibfnamefont{M.}~\bibnamefont{Horodecki}},
  \bibinfo{author}{\bibfnamefont{P.}~\bibnamefont{Horodecki}},
  \bibnamefont{and}
  \bibinfo{author}{\bibfnamefont{R.}~\bibnamefont{Horodecki}},
  \bibinfo{journal}{Phys. Lett. A} \textbf{\bibinfo{volume}{223}},
  \bibinfo{pages}{1} (\bibinfo{year}{1996}).

\bibitem[{\citenamefont{Augusiak et~al.}(2008)\citenamefont{Augusiak,
  Horodecki, and Demianowicz}}]{augusiak}
\bibinfo{author}{\bibfnamefont{R.}~\bibnamefont{Augusiak}},
  \bibinfo{author}{\bibfnamefont{R.}~\bibnamefont{Horodecki}},
  \bibnamefont{and}
  \bibinfo{author}{\bibfnamefont{M.}~\bibnamefont{Demianowicz}},
  \bibinfo{journal}{Phys. Rev.} \textbf{\bibinfo{volume}{77}},
  \bibinfo{pages}{030301(R)} (\bibinfo{year}{2008}).

\bibitem[{\citenamefont{Verstraete et~al.}(2001)\citenamefont{Verstraete,
  Audenaert, and DeMoor}}]{ver}
\bibinfo{author}{\bibfnamefont{F.}~\bibnamefont{Verstraete}},
  \bibinfo{author}{\bibfnamefont{K.}~\bibnamefont{Audenaert}},
  \bibnamefont{and} \bibinfo{author}{\bibfnamefont{B.}~\bibnamefont{DeMoor}},
  \bibinfo{journal}{Phys. Rev. A} \textbf{\bibinfo{volume}{64}},
  \bibinfo{pages}{012316} (\bibinfo{year}{2001}).

\bibitem[{\citenamefont{Bloore}(1976)}]{bloore}
\bibinfo{author}{\bibfnamefont{F.~J.} \bibnamefont{Bloore}},
  \bibinfo{journal}{J. Phys. A} \textbf{\bibinfo{volume}{9}},
  \bibinfo{pages}{2059} (\bibinfo{year}{1976}).

\bibitem[{\citenamefont{Joe}(2006)}]{joe}
\bibinfo{author}{\bibfnamefont{H.}~\bibnamefont{Joe}}, \bibinfo{journal}{J.
  Multiv. Anal.} \textbf{\bibinfo{volume}{97}}, \bibinfo{pages}{2177}
  (\bibinfo{year}{2006}).

\bibitem[{\citenamefont{Brown}(2001)}]{cylindrical}
\bibinfo{author}{\bibfnamefont{C.~W.} \bibnamefont{Brown}},
  \bibinfo{journal}{J. Symbolic Comput.} \textbf{\bibinfo{volume}{31}},
  \bibinfo{pages}{521} (\bibinfo{year}{2001}).

\bibitem[{\citenamefont{Strzebonski}(2002)}]{strzebonski}
\bibinfo{author}{\bibfnamefont{A.}~\bibnamefont{Strzebonski}},
  \bibinfo{journal}{Mathematica Journal} \textbf{\bibinfo{volume}{7}},
  \bibinfo{pages}{10} (\bibinfo{year}{2002}).

\bibitem[{\citenamefont{Szarek et~al.}(2006)\citenamefont{Szarek, Bengtsson,
  and {\.Z}yczkowski}}]{sbz}
\bibinfo{author}{\bibfnamefont{S.}~\bibnamefont{Szarek}},
  \bibinfo{author}{\bibfnamefont{I.}~\bibnamefont{Bengtsson}},
  \bibnamefont{and}
  \bibinfo{author}{\bibfnamefont{K.}~\bibnamefont{{\.Z}yczkowski}},
  \bibinfo{journal}{J. Phys. A} \textbf{\bibinfo{volume}{39}},
  \bibinfo{pages}{L119} (\bibinfo{year}{2006}).

\end{thebibliography}

\end{document}